\documentclass[aps,prl,reprint,twocolumn,superscriptaddress,preprintnumbers]{revtex4-1}

\usepackage{amsthm}
\usepackage{amsmath}
\usepackage{graphicx}
\usepackage{slashed}
\usepackage{amssymb}
\usepackage{float}
\usepackage[utf8]{inputenc}
\usepackage[T1]{fontenc}
\usepackage[colorlinks=True, citecolor=blue, urlcolor=blue, linkcolor=blue]{hyperref}

\newcommand*\diff{\mathop{}\!\mathrm{d}}

\newcommand{\nn}{\nonumber}
\newcommand{\h}[1]{\hat{#1}}

\newcommand{\be}{\begin{eqnarray}}
\newcommand{\ee}{\end{eqnarray}}

\newcommand{\ml}{\mathcal}
\newcommand{\bs}{\boldsymbol}
\newcommand{\Tr}{\mathrm{Tr}}

\begin{document}

\title{SU(2) Gauge Theory in $2+1$ Dimensions on a Plaquette Chain Obeys the Eigenstate Thermalization Hypothesis}

\author{Xiaojun Yao}
\email{xjyao@uw.edu}
\affiliation{InQubator for Quantum Simulation, Department of Physics, University of Washington, Seattle, Washington 98195, USA}

\preprint{IQuS@UW-21-047}
\begin{abstract}
We test the eigenstate thermalization hypothesis (ETH) for 2+1 dimensional SU(2) lattice gauge theory. By considering the theory on a chain of plaquettes and truncating basis states for link variables at $j=1/2$, we can map it onto a quantum spin chain with local interactions and numerically exactly diagonalize the Hamiltonian for reasonably large lattice sizes. We find energy level repulsion in momentum sectors with no remaining discrete symmetry. We study two local observables made up of Wilson loops and calculate their matrix elements in the energy eigenbasis, which are shown consistent with the ETH. 
\end{abstract}

\maketitle

{\it Introduction.} How an isolated quantum system thermalizes is a long standing question~\cite{d2016quantum,deutsch2018eigenstate,mori2018thermalization}. In particular, we want to know how expectation values of local observables and their fluctuations approach predictions from thermal statistics such as the microcanonical (mc) ensemble after the system is perturbed out of equilibrium. A significant progress in our understanding has been achieved over the last thirty years, highlighted in the formulation of the eigenstate thermalization hypothesis (ETH)~\cite{PhysRevA.43.2046,PhysRevE.50.888,rigol2008thermalization}. Many systems that are nonintegrable and/or classically chaotic have been shown to obey the ETH (see recent reviews~\cite{d2016quantum,deutsch2018eigenstate,mori2018thermalization}). Known exceptions of the ETH include integrable systems~\cite{rigol2007relaxation,calabrese2011quantum,khatami2013fluctuation}, many-body localizations~\cite{anderson1958absence,PhysRevLett.114.170505,nandkishore2015many,halimeh2021stabilizing} and quantum scars~\cite{bernien2017probing,shiraishi2017systematic,turner2018weak,chanda2020confinement,schecter2019weak,aramthottil2022scar,Banerjee:2020tgz,PhysRevLett.124.160604,PhysRevLett.127.150601}.

Although the ETH has been widely scrutinized in many quantum systems, very few studies tested the ETH for gauge theories (here we focus on relativistic quantum field theories invariant under local gauge transformations given by Lie groups). Testing the ETH for gauge theories is not just an academic question, but has many practical applications in understanding how systems of Standard Model particles thermalize. One example is the reheating stage of the early Universe right after the inflation~\cite{PhysRevD.23.347}, at the end of which the Universe reaches a radiation-dominated thermal equilibrium~\cite{Kofman:1997yn,Allahverdi:2010xz,Amin:2014eta,Nguyen:2019kbm,vandeVis:2020qcp,McDonough:2020tqq}. Another example is the initial stage of relativistic heavy ion collisions, when highly occupied gluon states isotropize and reach local equilibrium approximately~\cite{Muller:2011ra,Berges:2020fwq} so that viscous hydrodynamics can be applied to describe the following evolution, which is critical in our understanding of the thermal behaviors in various particles' yields~\cite{Kolb:2003dz,stachel2014confronting}. Many studies have been devoted to understand the initial rapid thermalization in heavy ion collisions, by using techniques such as perturbative calculations~\cite{Baier:2000sb,Kurkela:2018wud}, the color-glass condensate framework~\cite{McLerran:1993ni,McLerran:1993ka,Muller:2005yu,Fries:2008vp,Dusling:2010rm,Schenke:2012wb} and the AdS/CFT correspondence~\cite{Chesler:2008hg,Chesler:2009cy,Balasubramanian:2010ce,Heller:2012km,vanderSchee:2012qj,Balasubramanian:2013rva,Balasubramanian:2013oga}, highlighted by the recent discoveries of attractor behaviors in certain quantities~\cite{Heller:2011ju,Heller:2015dha,Heller:2016rtz,Romatschke:2017vte,Strickland:2017kux,Blaizot:2017ucy,Strickland:2018ayk,Behtash:2019txb,Giacalone:2019ldn,Brewer:2019oha,Brewer:2022vkq}, which are determined by slow modes that govern the early time dynamics before hydrodynamics becomes applicable.

Despite the many achievements in our understanding of thermalization in the early Universe and heavy ion collisions, an explanation fully based on the quantum wavefunction is still desirable, since it can provide insights about thermalization from a different perspective. The ETH is one possible quantum explanation in which the quantum wavefunction {\it does not decohere} during the thermalization process. Previous studies have shown that non-Abelian gauge theories in 3+1 dimensions are classically chaotic~\cite{Muller:1992iw,Biro:1994sh,Heinz:1996wx,Bolte:1999th}, which implies that the ETH is very likely to hold for them.
However, there is no direct and explicit demonstration of the ETH for non-Abelian gauge theories. One difficulty is the rapid growth of the Hilbert space as the system size on a lattice increases and/or any truncation is removed, which prohibits exact diagonalization of the Hamiltonian. 

Here in this letter, we provide the first test of the ETH for non-Abelian gauge theories. Motivated by recent developments of quantum simulation for gauge theories~\cite{PhysRevLett.110.125303,tagliacozzo2013simulation,Klco:2018kyo,Kaplan:2018vnj,Raychowdhury:2018osk,rico2018so,Klco:2019evd,Raychowdhury:2019iki,Davoudi:2020yln,Shaw:2020udc,celi2020emerging,Ciavarella:2021nmj,Ciavarella:2021lel,deJong:2021wsd,Kan:2021nyu,Nguyen:2021hyk,Bauer:2021gek,Farrell:2022wyt,Farrell:2022vyh,Yao:2022eqm,Ciavarella:2022qdx,Davoudi:2022uzo,Mueller:2022xbg,Grabowska:2022uos,Kane:2022ejm,Funcke:2022uwc,Angelides:2022pah,Davoudi:2022xmb,Kadam:2022ipf,Florio:2023dke,Funcke:2023jbq}, we use the Kogut-Susskind Hamiltonian formulation for non-Abelian gauge theories~\cite{PhysRevD.11.395}. More specifically, we consider 2+1 dimensional SU(2) gauge theory on a chain of plaquettes. By truncating the physical Hilbert space, we are able to exactly diagonalize the Hamiltonian on lattices of various sizes and investigate the asymptotic scaling properties of local observables as the system size increases, which is the essential ingredient of the ETH. 

{\it Review of ETH.} We consider the time evolution of an isolated quantum system with an initial state $\rho(t=0)$. At a later time $t$, the expectation value of a local observable $O$ is given by
\begin{align}
\label{eqn:O_t}
\langle O \rangle (t) = \Tr[O\rho(t)] = \sum_{n,m} O_{nm} \rho_{mn}(0) e^{i(E_n-E_m)t} \,,
\end{align}
where $O_{nm} = \langle n| O | m \rangle $, $\rho_{mn} = \langle m| \rho | n \rangle $ and $|n\rangle$ denotes eigenstates of $H$ with eigenenergies $E_n$. The question of interest is how Eq.~\eqref{eqn:O_t} approaches the thermal equilibrium value, e.g., the microcanonical ensemble average $\langle O \rangle_{\rm mc}(E)$ where the system's energy is fixed by $E=\Tr(H\rho)$.

The question can be answered by the ETH, which states that matrix elements of the observable in the energy eigenbasis are given by~\cite{d2016quantum,deutsch2018eigenstate,mori2018thermalization}
\begin{align}
\label{eqn:eth}
O_{nm} = \langle O \rangle_{\rm mc}(E) \delta_{nm} + e^{-S(E)/2}f(E,\omega) R_{nm}\,,
\end{align}
where $E=(E_n+E_m)/2$, $\omega = E_n-E_m$, $S(E)$ denotes the thermodynamic entropy of the system at the energy $E$, which scales as the system size, and $R_{nm}$ is a random variable with zero mean and unit variance. The function $f(E,\omega)$ is smooth and gradually vanishes as $\omega\to\infty$. Now we are going to use the assumption in Eq.~\eqref{eqn:eth} to explain how the observable expectation value in Eq.~\eqref{eqn:O_t} approaches the thermal equilibrium value after some time. To this end, we consider a generic initial state that has significant overlaps with $N_{s}$ eigenstates. Typically we expect $\rho_{mn}(0) \sim {1}/{N_s}$ if $m,n$ are among these $N_{s}$ eigenstates and $\rho_{mn}(0) \sim 0$ otherwise~\cite{rigol2008thermalization}. At time $t=0$, the off-diagonal contribution to Eq.~\eqref{eqn:O_t} is given by
\begin{align}
\label{eqn:off_t=0}
\sum_{n,m}^{n\neq m} O_{nm} \rho_{mn}(0) \sim \frac{N^2_s}{N_s} O_{\rm off{\text -}diag}^{\rm typical}  = N_s O_{\rm off{\text -}diag}^{\rm typical} \,,
\end{align}
where $O_{\rm off{\text -}diag}^{\rm typical}$ is the typical value of the off-diagonal matrix elements of the operator $O$ in the basis of those $N_s$ eigenstates. On the other hand, at large time $t$, the off-diagonal contribution becomes
\begin{align}
\label{eqn:off_large_t}
\sum_{n,m}^{n\neq m} e^{i(E_n-E_m)t} O_{nm} \rho_{mn}(0) \sim \frac{\sqrt{N^2_s}}{N_s} O_{\rm off{\text -}diag}^{\rm typical}  = O_{\rm off{\text -}diag}^{\rm typical} \,,
\end{align}
where the number of contributing off-diagonal terms is the square root of that at $t=0$ due to the dephasing at large time $t$~\cite{rigol2008thermalization}. Because of the exponential decay factor in the off-diagonal part of Eq.~\eqref{eqn:eth} with respect to the diagonal part, the off-diagonal contribution to $\langle O \rangle(t)$ at large time $t$ is much smaller than the diagonal microcanonical contribution, which means the observable approximately reaches its thermal expectation value $\langle O \rangle_{\rm mc}$. The time scale at which the transition from Eq.~\eqref{eqn:off_t=0} to Eq.~\eqref{eqn:off_large_t} happens gives the thermalization time scale.

In the following, we will test if Eq.~\eqref{eqn:eth} holds for 2+1 dimensional SU(2) lattice gauge theory.

{\it 2+1 dimensional SU(2) lattice gauge theory.} The Kogut-Susskind Hamiltonian of the theory can be written as~\cite{PhysRevD.11.395} (see also Ref.~\cite{sm})
\begin{align}
\label{eqn:H}
H = \frac{g^2}{2}\sum_{\rm links} (E_i^a)^2 - \frac{2}{a^2g^2} \sum_{\rm plaquettes} Z({\bs n}) \,,
\end{align}
where $a$ in the denominator is the lattice spacing and that in the superscript denotes SU(2) indexes that are implicitly summed over, $g$ is the gauge coupling with the mass dimension $[g]=0.5$ in 2+1 dimensions, $i=x$ or $y$ for spatial directions (implicitly summed), ${\bs n} = (n_x, n_y)$ represents a lattice point, and $Z({\bs n})$ is the plaquette operator defined as
\begin{align}
Z({\bs n}) &= \Tr[U^\dagger({\bs n},\hat{y}) U^\dagger({\bs n}+\hat{y},\hat{x}) U({\bs n}+\hat{x},\hat{y}) U({\bs n},\hat{x})] \,,\nn\\
U({\bs n},\hat{i}) &= e^{ia A_i^a({\bs n})T^a} \,,
\end{align}
where $U({\bs n},\hat{i})$ is a link variable on the link from ${\bs n}$ to ${\bs n}+\h{i}$ and $T^a=\sigma^a/2$ is the generator of the SU(2) group in the fundamental representation. 
The electric field operators $E_i^a$ in Eq.~\eqref{eqn:H} can generate a gauge transformation either on the left end of a link (denoted as $E_{Li}^a$) or on the right end (labeled as $E_{Ri}^a$) and satisfy the following commutation relations~\cite{Zohar:2014qma}
\begin{align}
\label{eqn:commutators}
&[ E_{Li}^a({\bs n}+\hat{i}/2), U({\bs n},\hat{j}) ] = -\delta_{ij} T^a  U({\bs n},\hat{j}) \,,\nn\\
&[ E_{Ri}^a({\bs n}+\hat{i}/2), U({\bs n},\hat{j}) ] = \delta_{ij}  U({\bs n},\hat{j}) T^a \,,\nn\\
&[E_{Li}^a, E_{Li}^b] = if^{abc} E_{Li}^c\,,\quad [E_{Ri}^a, E_{Ri}^b] = if^{abc} E_{Ri}^c \,,
\end{align}
where the argument ${\bs n}+\hat{i}/2$ of the electric fields means they live on the link between ${\bs n}$ and ${\bs n}+\h{i}$ and $f^{abc} = \varepsilon^{abc}$ is the structure constant of the SU(2) group. 

Since electric fields can generate gauge transformations on both the left and right hand sides of a link variable, the link variable can be represented by two irreducible representations with the same highest weight (two angular momentum states labeled by $|jm\rangle$ with the same $j$), i.e., $|jm_Lm_R\rangle$. They serve as basis states in the Hilbert space and are normalized as $\langle j'm_L'm_R'|jm_Lm_R\rangle = \delta_{j'j}\delta_{m_L'm_L}\delta_{m_R'm_R}$. In this basis, the matrix element of a link variable $U_{n_Ln_R}$ is ($U$ is a SU(2) matrix in the fundamental representation and $U_{n_Ln_R}$ is one entry with $n_L,n_R\in\{1/2,-1/2\}$)~\cite{Byrnes:2005qx,Zohar:2014qma}
\begin{align}
\label{eqn:matrix_U}
&\langle j'm_L'm_R' | U_{n_Ln_R} | j m_L m_R \rangle = \sqrt{({2j+1})/({2j'+1})} \nn\\
&  \qquad \times \langle j'\, m_L'| j \, m_L; 1/2\, n_L \rangle \langle j\,  m_R; 1/2\, n_R | j'\,  m_R' \rangle \,,
\end{align}
where $\langle j'\, m'| j \, m; J\, M \rangle $ denotes Clebsch-Gordan coefficients. The matrix representation for $U^\dagger$ can be obtained by taking Hermitian conjugate of Eq.~\eqref{eqn:matrix_U}. The matrix element for the electric part of the Hamiltonian is diagonal
\begin{align}
\sum_{\rm links}(E_i^a)^2 |jm_Lm_R\rangle = j(j+1)|jm_Lm_R\rangle \,.
\end{align}

Only states satisfying the Gauss's law are physical states. The Gauss's law at each lattice site ${\bs n}$ can be written as
\begin{align}
\sum_{i=x,y} E_{Li}^a({\bs n}+\hat{i}/2) + \sum_{i=x,y} E_{Ri}^a({\bs n}-\hat{i}/2) = 0\,.
\end{align}
Physically, it means all the link variables joining the same lattice site transform as a singlet together for physical states, i.e., they are invariant under local gauge transformations.

\begin{figure}
\centering
\includegraphics[width=0.45\textwidth]{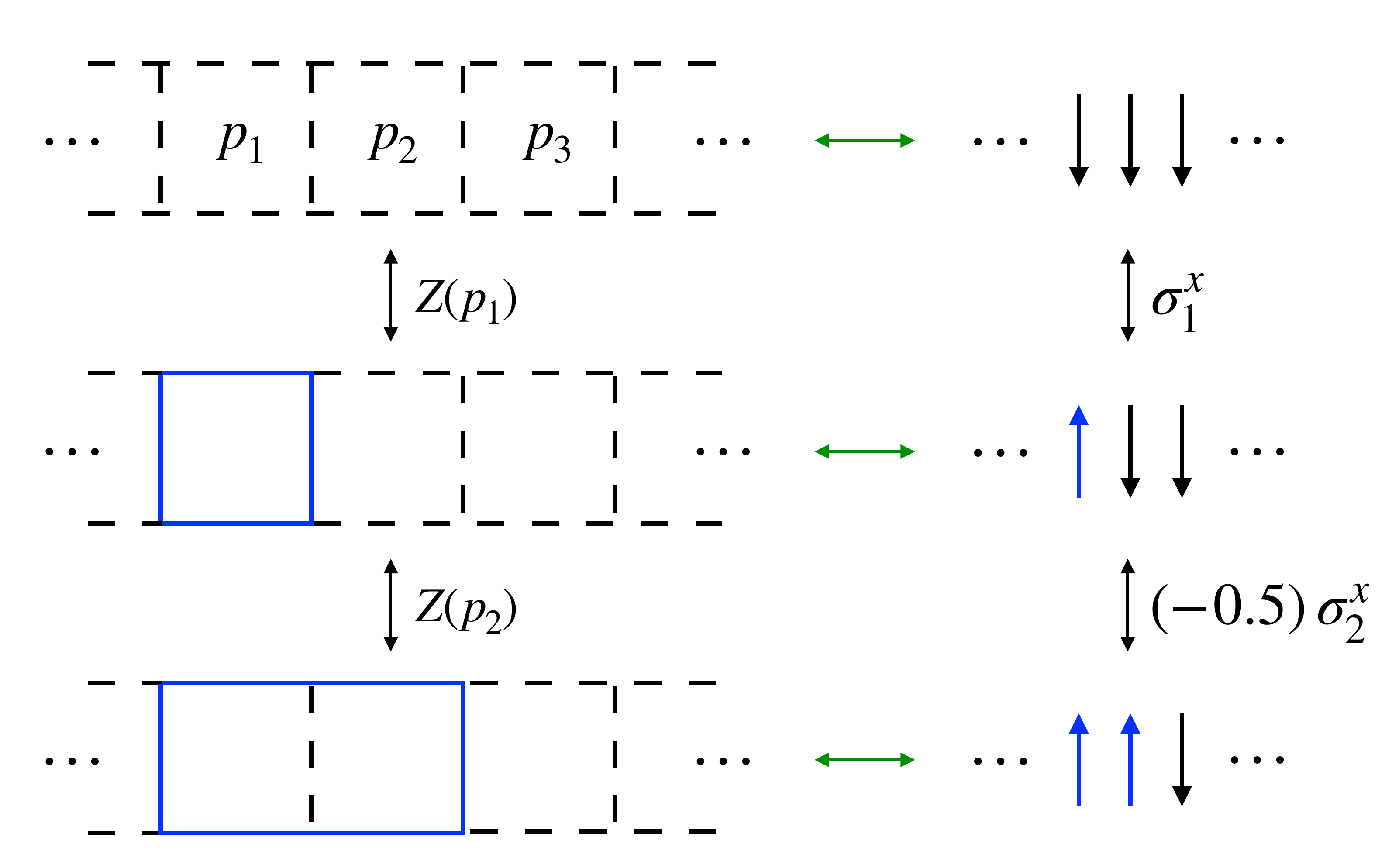}
\caption{The bijective map between the 2+1 dimensional SU(2) lattice gauge theory on a plaquette chain with the basis truncated at $j=1/2$ and a quantum spin chain. The three plaquettes ($p_1,p_2,p_3$) on the left are mapped onto three spins on the right. Black dashed lines on the left represent $j=0$ link states while blue solid lines mean $j=1/2$ link states. The plaquette operator $Z(p_i)$ corresponds to $\sigma^x_{i}$, up to a prefactor determined by the two nearest neighbors.}
\label{fig:map}
\end{figure}

{\it Map onto spin chain.} We consider a Hilbert space with a truncation at $j=1/2$, which is valid in the strong coupling (infrared) limit. Going beyond the $j=1/2$ truncation is left for future studies. We study physical states living on a chain of plaquettes with periodic boundary conditions, shown in Fig.~\ref{fig:map}. We only consider states that are generated by acting plaquette operators on the bare vacuum, which do not have topologically nontrivial gauge flux around the chain. A plaquette operator $Z({\bs n})$ acting on a state with four edges being in the $j=0$ state creates a state with four edges in the $j=1/2$ states that form linear combinations to transform as SU(2) singlets at each corner. Acting another $Z({\bs n})$ operator on the same plaquette creates either a singlet state with four edges in the $j=0$ state or $j=1$ states, the latter of which are neglected due to the truncation. When two adjacent plaquette operators $Z({\bs n})Z({\bs n}+\hat{x})$ act on a state with all the relevant links being in $j=0$, two physical states are generated. In one state, the overlapped link (the common edge shared by the two plaquettes) is in the $j=0$ state while in the other, the overlapped link is in the $j=1$ states. We only keep the former state in our current study for a consistent truncation. If we represent a plaquette state with four edges being in the $j=0$ ($j=1/2$) state as a spin-down (spin-up) state, the plaquette operator $Z({\bs n})$ can be represented as a Pauli matrix $\sigma^x_{\bs n}$, up to a prefactor determined by the two nearest neighbors, which will be shown below. Furthermore, a plaquette state with four edges in the $j=0$ state contributes zero to the electric part of the Hamiltonian, while an isolated (neighboring links are in the $j=0$ states) plaquette state with four edges in the $j=1/2$ states contributes $\frac{g^2}{2}\cdot\frac{3}{4}\cdot 4$ to the electric part of the Hamiltonian. If two neighboring plaquettes are both in the $j=1/2$ states, we need to subtract $\frac{g^2}{2}\cdot\frac{3}{4}\cdot 2$ from their contribution to the electric part, since the overlapped link is in the $j=0$ state and contributes vanishingly. Putting all these together, we find the Hamiltonian of the SU(2) gauge theory on a plaquette chain with a basis truncated at $j=1/2$ can be mapped onto a quantum spin chain, shown in Fig.~\ref{fig:map}
\begin{align}
H = &\ \frac{3}{2}g^2\sum_{i=0}^{N-1}\frac{\sigma_i^z+1}{2} - \frac{3}{4}g^2\sum_{i=0}^{N-1}\frac{\sigma_i^z+1}{2}\frac{\sigma_{i+1}^z+1}{2} \nn \\
&  - \frac{2}{a^2g^2} \sum_{i=0}^{N-1}
\big(-0.5\big)^{\frac{\sigma_{i-1}^z+\sigma_{i+1}^z+ 2}{2}}
\sigma_i^x \,.
\end{align}
Up to an irrelevant constant, this Hamiltonian can be rewritten as (see Ref.~\cite{Hayata:2021kcp,ARahman:2022tkr} for a similar expression)
\begin{align}
\label{eqn:H_ising}
aH =&\ J \sum_{i=0}^{N-1}\sigma_i^z\sigma_{i+1}^z + h_z \sum_{i=0}^{N-1}\sigma_i^z \nn \\
&+ h_x \sum_{i=0}^{N-1} \frac{1-3\sigma_{i-1}^z}{4} \frac{1-3\sigma_{i+1}^z}{4} \sigma_i^x \,, 
\end{align}
where $J = -3ag^2/16$, $h_z=3ag^2/8$ and $h_x = -2/(ag^2)$. Under the periodic boundary condition, $\sigma_N^i=\sigma_0^i$. The Hamiltonian is rescaled to be unitless and so are the parameters $J$, $h_z$ and $h_x$. This Hamiltonian is similar to the quantum Ising chain with a transverse field that is known to be nonintegrable and exhibit ETH behaviors~\cite{kim2013ballistic,PhysRevE.90.052105}, but the $\sigma_i^x$ term here is different. The difference is a result of the Gauss's law and the Clebsch-Gordan coefficients in the matrix elements of link variables shown in Eq.~\eqref{eqn:matrix_U} and can be obtained from e.g., Eq.~(32) of Ref.~\cite{sm}. We expect the spin model shown in Eq.~\eqref{eqn:H_ising} to be nonintegrable.

For reasonably large values of $N$, we are able to numerically exactly diagonalize the Hamiltonian in Eq.~\eqref{eqn:H_ising} by using symmetries of the system to reduce the Hilbert space size. One symmetry is translational invariance $[H,\hat{T}] = 0$, where $\hat{T}$ denotes a translation operator by one lattice site. Thus, we can simultaneously diagonalize $H$ and $\hat{T}$. The eigenstates of $\hat{T}$ are momentum states $|k_i\rangle$ ($k_i=2\pi i/N$, $i=0,1,\cdots,N-1$), which can be constructed easily~\cite{sandvik2010computational} (see also Ref.~\cite{sm}). We can then construct the Hamiltonian in each momentum sector and diagonalize therein. The Hamiltonian is block-diagonal with vanishing off-diagonal elements between different momentum sectors. Furthermore, the Hamiltonian is invariant under reflection $i\to N-i$ (parity): The $k=0$ and $k=\pi$ sectors are invariant while the $k_i$ sector turns to the $k_{N-i}$ sector.
So we will only study the momentum sectors $k_i$ up to $i=\lfloor N/2 \rfloor$.

\begin{figure}
\centering
\includegraphics[width=0.45\textwidth]{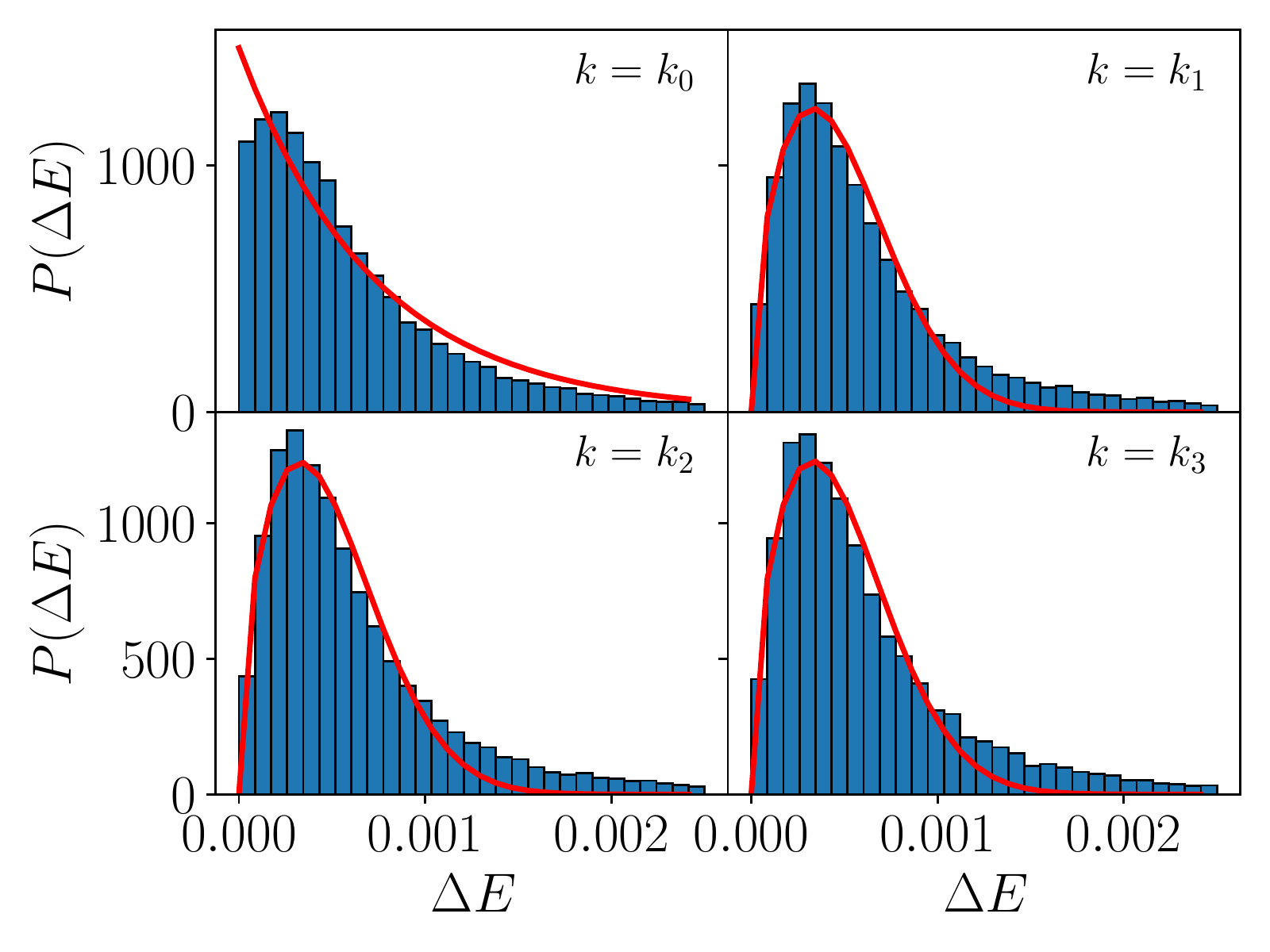}
\caption{Distributions of energy level spacing $\Delta E$ in $k_i$ ($i=0,1,2,3$) sectors for $N=19$. The distributions in the nonzero momentum sectors exhibit the Wigner-Dyson statistics with level repulsion while the distribution in the $k_0$ sector is closer to the Poisson form due to the remaining parity symmetry.}
\label{fig:wd}
\end{figure}

{\it Results.} 
We need to choose parameter values in numerical studies. In principle, the coupling $g$ is a function of $a$ determined by the renormalization group equation that is obtained by requiring physical observables are independent of $a$ in the limit $a\to0$. Here we do not aim at extracting physical quantities out of the calculation but just want to test the ETH, so picking up one value for $ag^2$ suffices as long as the Hamiltonian~\eqref{eqn:H_ising} is not integrable. We choose $ag^2=1.2$ so that the ETH scaling can be manifest in finite systems that are numerically accessible.

First we study the statistics of energy level spacing. We list all energy eigenvalues in each momentum sector in an ascending order and calculate their nearest gaps $\Delta E = E_{n+1} - E_n$. The distributions of $\Delta E$ in the first four momentum sectors $k_i$, $i=0,1,2,3$ for $N=19$ are shown in Fig.~\ref{fig:wd}, where each momentum sector contains 27594 states (the $k=0$ sector has two more). In the sectors with nonzero momenta, the distribution resembles the Wigner-Dyson statistics, featured in the level repulsion (the distribution vanishes at $\Delta E=0$). The Wigner-Dyson statistics is often found in systems that are nonintegrable and chaotic classically~\cite{bohigas1984characterization}. The red lines shown in the nonzero momenta cases are fits from the Wigner surmise $P_{\rm ws}(\Delta E) = a(\Delta E)^b \exp[-c (\Delta E)^2]$ with $a,b,c$ parameters. The fit in the tail region can be much improved if only the middle part of the eigenenergy spectrum is used, as shown in Ref.~\cite{sm}. The zero momentum sector is special here: There is no level repulsion and the distribution is more similar to the Poisson statistics rather than the Wigner-Dyson one. The red curve is a fit from the Poisson statistics of the form $P_{\rm p}(\Delta E) = a \exp(-b\Delta E)$ with $a,b$ parameters differing from those in the Wigner surmise. The absence of level repulsion in the zero momentum sector is caused by the remaining parity symmetry mentioned earlier (level statistics in each parity sector can be found in Ref.~\cite{sm}, as well as fitted parameter values). It is known that discrete symmetries can invalidate the Wigner-Dyson statistics of level separations. However, the ETH, which is a statement about the eigenstates, is still expected to hold, even in the presence of discrete symmetries~\cite{santos2010localization}. In the following results, we will include all momentum $k_i$ sectors from $i=0$ to $\lfloor N/2 \rfloor$.

\begin{figure}
\centering
\includegraphics[width=0.45\textwidth]{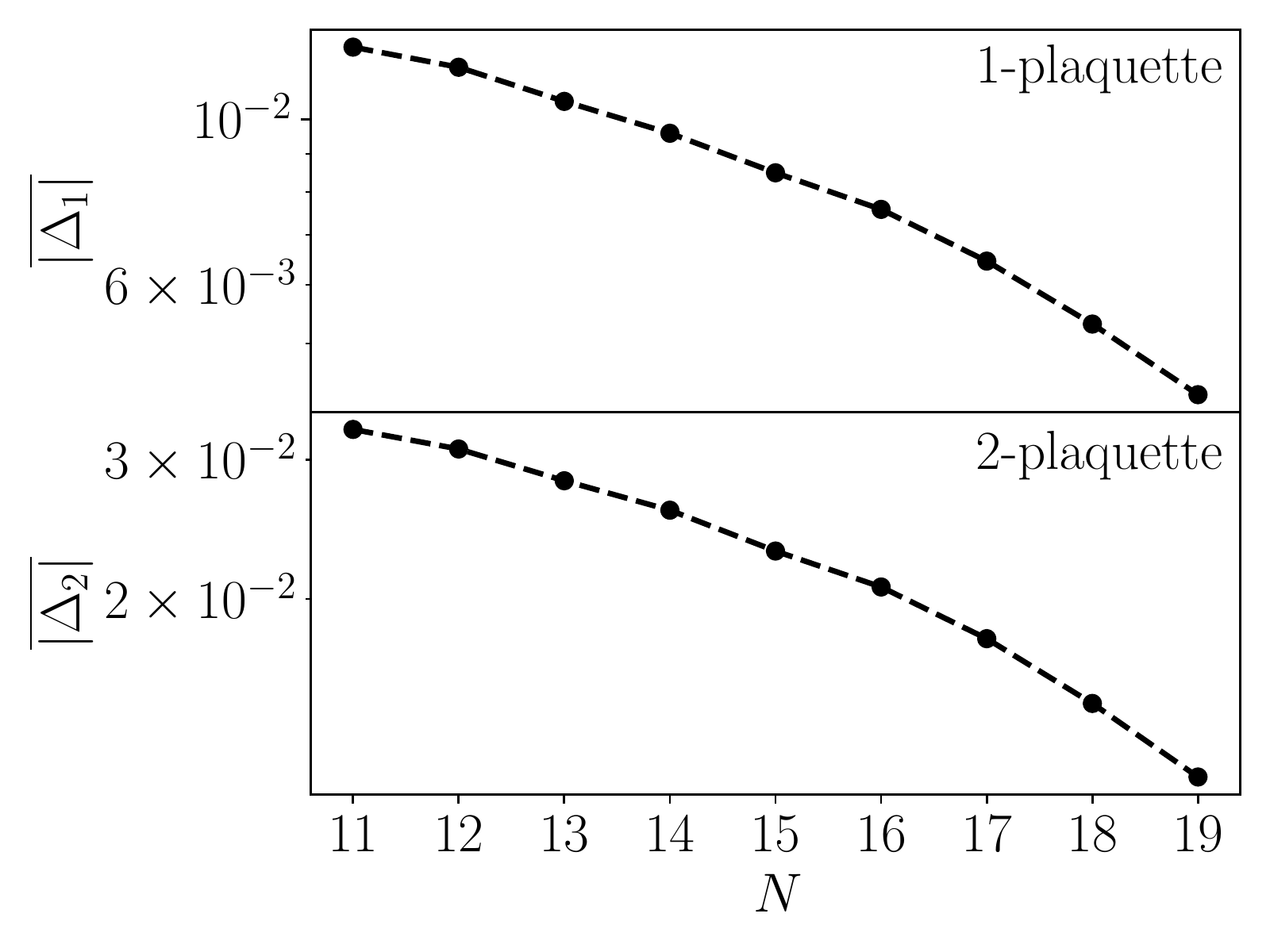}
\caption{Averaged magnitude of the difference between the operator expectation value and the microcanonical ensemble proxy as a function of the system size.}
\label{fig:diag}
\end{figure}

Next we test the diagonal part of the ETH. The most crucial aspect to demonstrate is the exponential decrease of the second term in Eq.~\eqref{eqn:eth} with the system size (the entropy is proportional to the system size $S\propto N$). The two local observables we study are 1-plaquette ($O_1$) and 2-plaquette ($O_2$) operators, which correspond to square and rectangular Wilson loops. For an eigenstate $|n\rangle$ with an energy $E_n$, we consider its nearest 20 neighbors in energies (10 above and 10 below). We use their average as a proxy for the microcanonical ensemble average at the same energy. Then we compute the difference between the expectation value of an operator in the eigenstate $|n\rangle$ and its microcanonical ensemble proxy
\begin{align}
\label{eqn:delta_n}
\Delta_i(n) = \langle n | O_i | n\rangle - \frac{1}{21}\sum_{m=n-10}^{n+10} \langle m | O_i | m \rangle \,.
\end{align}
If the ETH holds, we will expect the average value of $|\Delta_i(n)|$, i.e., $\overline{|\Delta_i|}$ to decrease exponentially with the system size $N$. (All states are used in the calculation of the average except for the 10 lowest and 10 highest eigenenergy states.) Fig.~\ref{fig:diag} clearly shows this exponential decrease and thus demonstrating the diagonal part of the ETH for the majority of states. In fact, Fig.~\ref{fig:diag} seems to suggest the decrease is faster than an exponential in $N$. However, if we only use the middle two thirds of the eigenstates (ordered by their eigenenergies) to calculate the average, the $N=16,17,18,19$ points exhibit a better agreement with an exponential decrease in $N$~\cite{sm}.

\begin{figure}
\centering
\includegraphics[width=0.45\textwidth]{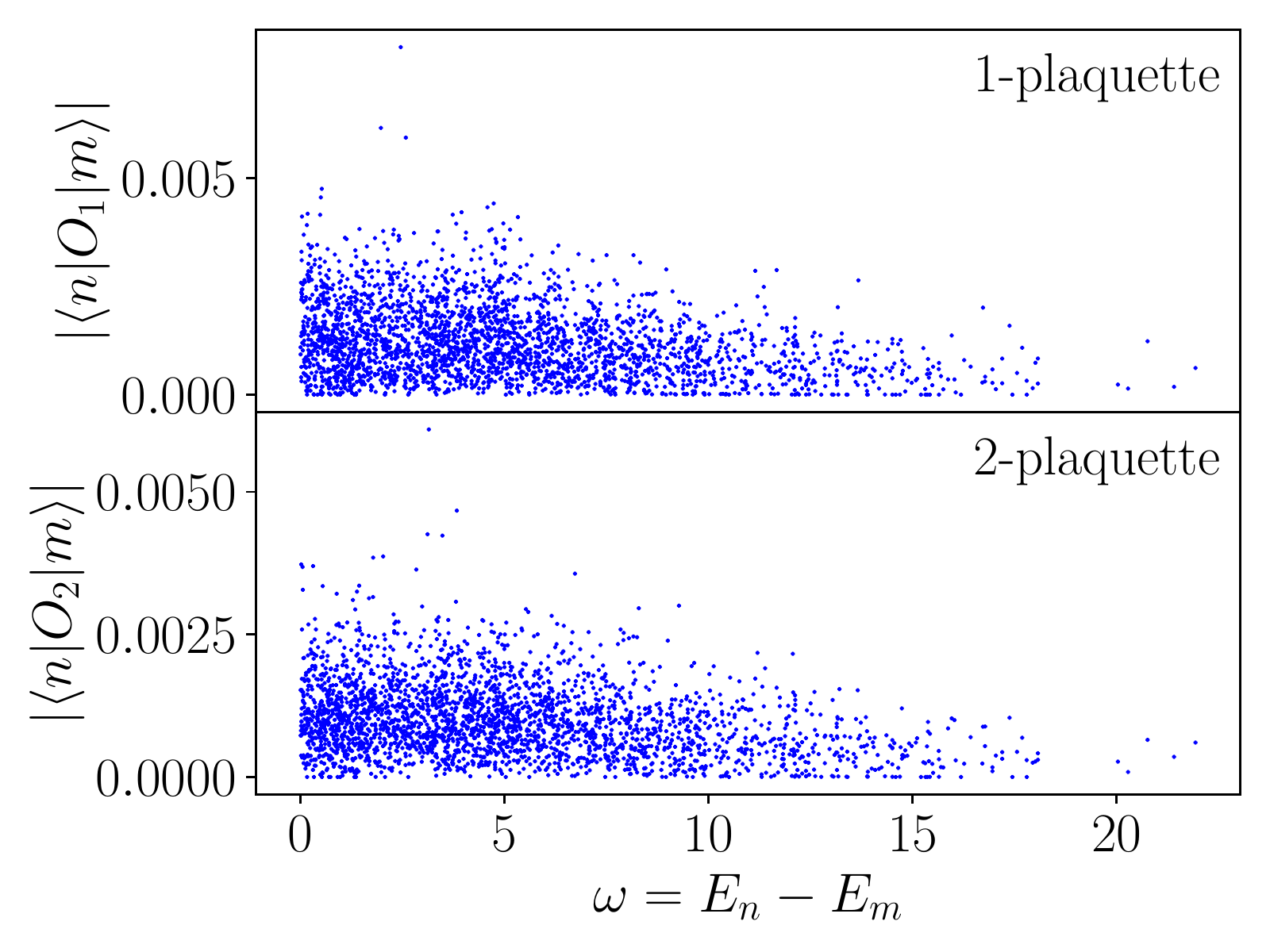}
\caption{Magnitudes of off-diagonal matrix elements $|\langle m | O_i | n \rangle|$ decrease as a function of $\omega$ for the two operators in the $N=17$ case.}
\label{fig:off}
\end{figure}

Finally, we study the off-diagonal part of the ETH. The most important thing to show is the function $f(E,\omega)$ vanishing at large $\omega$. To this end, we study all pairs of states $|n\rangle$ and $|m\rangle$ whose total energy $(E_n+E_m)/2$ falls between $E-\epsilon$ and $E+\epsilon$. We calculate the absolute value of the matrix element $|\langle m | O_i | n \rangle|$ as a function of $\omega = E_n-E_m$ (we choose $E_n>E_m$ without loss of generality). We choose $E=1$ and $\epsilon=10^{-5}$ and plot the result for the $N=17$ case in Fig.~\ref{fig:off}, which explicitly displays of the decrease of $f(E,\omega)$ towards zero as $\omega$ increases. For $N=17$, the lowest and highest eigenenergies are roughly $-24.76$ and $16.69$ respectively. So the number of terms with $\omega\gtrsim20$ in the figure turns to zero.

{\it Conclusions.} In this letter, we tested the ETH for 2+1 dimensional SU(2) lattice gauge theory on a chain of plaquettes with a truncation at $j=1/2$ in the electric basis. In this simple setup, we mapped the Hamiltonian of the SU(2) gauge theory onto a quantum spin chain with local interactions. By exact diagonalization, we studied the statistics of level separations and showed level repulsion in momentum sectors that have no reflection symmetry. Furthermore, we calculated matrix elements of local observables (Wilson loops) in the energy eigenbasis and demonstrated the scaling properties with the system size for both the diagonal and off-diagonal parts of the ETH. The simple Hamiltonian considered here can be easily studied on quantum hardwares such as the quantum annealer~\cite{PhysRevD.104.034501,dwave}, cold atoms~\cite{PhysRevLett.74.4091,PhysRevLett.110.125304,endres2016atom,ebadi2021quantum,quera}, trapped ions~\cite{haffner2008quantum,barreiro2011open,davoudi2020towards,quantinuum,ionq} and superconducting qubits~\cite{kjaergaard2020superconducting,imb,google}. Studies of SU(2) and SU(3) gauge theories on small lattices have been performed on IBM's quantum hardware~\cite{Klco:2019evd,Ciavarella:2021nmj,Ciavarella:2021lel,ARahman:2022tkr}.

Future studies should investigate cases with a plane or a volume of plaquettes, where the simple map used here does not work due to the Mandelstam constraint~\cite{mandelstam1968feynman,mandelstam1979charge,giles1981reconstruction}. One should also study cases with $j$ truncated at higher values, the SU(3) case and cases with fermions included. The Hamiltonian in these cases may not be easy to exactly diagonalize. But one may still be able to use quantum computers to simulate the time evolution and study various thermalization processes to test features of the ETH. Other interesting questions are whether quantum scars exist in these more general cases, entanglement Hamiltonian~\cite{Mueller:2021gxd} and non-Abelian ETH~\cite{Murthy:2022dao}. All these studies will deepen our understanding of thermalization in systems consisting of Standard Model particles, such as the early Universe and high energy nuclear collisions.

\begin{acknowledgments}
We would like to thank Anthony Ciavarella, Masanori Hanada, Marc Illa Subina, David Kaiser, Bruno Scheihing-Hitschfeld, Hersh Singh and Francesco Turro for useful discussions. We would also like to thank Berndt M\"uller and Martin Savage for comments on the draft.
This work was supported by the U.S. Department of Energy, Office of Science, Office of Nuclear Physics, InQubator for Quantum Simulation (IQuS) under Award Number DOE (NP) Award DE-SC0020970. This work was facilitated through the use of advanced computational, storage, and networking infrastructure provided by the Hyak supercomputer system at the University of Washington.
\end{acknowledgments}

\bibliography{main.bib}

\begin{widetext}
\newpage
\section{Supplemental Material}
\subsection{The Kogut-Susskind Hamiltonian of SU($N_c$) Pure Gauge Theory from the Continuum Lagrangian}
Here we focus on the case of 3 spatial dimensions. The 2D and 1D cases can be similarly worked out. The Lagrangian density of 3+1 dimensional SU($N_c$) non-Abelian gauge theory can be written as
\begin{align}
\ml{L} = -\frac{1}{4g^2} F^{\mu\nu a}F_{\mu\nu}^a \,,
\end{align}
where the prefactor is no longer the standard $-1/4$ since we have redefined the gauge fields via $A^{\mu a} \to g A^{\mu a}$. In our notation, $\mu,\nu,\cdots$ denote Minkowski indexes while $i,j,k,\cdots$ represent spatial Euclidean indexes. The field strength is given by $F_{\mu\nu}^a = \partial_\mu A_\nu^a - \partial_\nu A_\mu^a + f^{abc} A_\mu^b A_\nu^c$. Here $f^{abc}$ denotes the structure constant of the SU($N_c$) group and is defined by $[T^a, T^b] = if^{abc} T_c$ where $T^a$ is the generator of the group, normalized as $\Tr(T^aT^b) = \delta^{ab}/2$. The canonical momentum conjugated to the field variable $A^{\mu a}$ is
\begin{align}
\Pi_\mu^a = \frac{\partial\ml{L}}{\partial(\partial^0A^{\mu a})} = -\frac{1}{g^2} F_{0\mu}^a \,.
\end{align}
The Euler-Lagrange equation of motion is
\begin{align}
(D^\mu F_{\mu\nu})^a = \partial^\mu F_{\mu\nu}^a + f^{abc} A^{\mu b} F_{\mu\nu}^c = 0 \,,
\end{align}
where $D^\mu$ is the covariant derivative in the adjoint representation.

The system has a constraint $\Pi_0^a = 0$. To obtain the Hamiltonian, we use axial gauge $A^{0a} = 0$ and impose the Gauss's law $D^\mu F_{\mu 0} = -D_i F_{i0} =0$ on states $|A^{i a}({\bs x})\rangle$ (the bold symbol indicates a Euclidean 3-vector). The state is specified by the field values at each spatial position ${\bs x}$ for each spatial direction $i$ and group index $a$. Only states satisfying the Gauss's law are physical states and appear in the physical Hilbert space:
\begin{align}
[D_i E_i({\bs x})]^a | A^{\mu b}({\bs y})\rangle = 0 \,,
\end{align}
at all spatial positions ${\bs x}$ for each $a$, where we have defined the non-Abelian electric field $E_i^a \equiv F_{0i}^a /g^2$. The canonical commutation relation between dynamical variables at equal time is given by
\begin{align}
\label{sm:EA_continuum}
[E_i^a({\bs x}), A_j^b({\bs y})] = i\delta_{ij}\delta^{ab} \delta^3({\bs x}-{\bs y}) \,.
\end{align}
The Hamiltonian density is then obtained as
\begin{align}
\label{sm:H_continuum}
\ml{H} = \frac{g^2}{2} (E_i^a)^2 + \frac{1}{4g^2}F_{ij}^aF_{ij}^a \,.
\end{align}
Furthermore, we can show the operator $[D_i E_i({\bs x})]^a$ is the generator of gauge transformation:
\begin{align}
\label{sm:DE_generator}
\exp\Big\{i \! \int \!\diff^3y \phi^b({\bs y}) [D_jE_j({\bs y})]^b \Big\} A_i^a({\bs x}) \exp\Big\{ \! - \! i \! \int \! \diff^3y \phi^b({\bs y}) [D_jE_j({\bs y})]^b \Big\} = A_i^a({\bs x}) + \partial_i \phi^a({\bs x}) - f^{abc} \phi^b({\bs x}) A_i^c({\bs x}) + \ml{O}(\phi^2)\,,
\end{align}
of which the right hand side is an infinitesimal local gauge transformation of the gauge field $A_i^a({\bs x})$ parametrized by $\phi^b({\bs x})$. Finally, for later convenience we define a spatial Wilson line in the fundamental representation as
\begin{align}
U({\bs y}, {\bs x}) = \ml{P} \exp\Big[ i \int_{\bs x}^{\bs y} \diff z_i A_i({\bs z}) \Big] \,,
\end{align}
where ${\bs A} = {\bs A}^aT^a $ and the symbol $\ml{P}$ denotes path ordering for a straight line from ${\bs x}$ to ${\bs y}$ parametrized by ${\bs z}$. One can show
\begin{align}
\label{sm:commutator_continuum}
[E_j^a(y_i), U(y_i, x_i)] &= - \delta_{ij} \delta^2({\bs 0}) T^a U(y_i, x_i) \,,\nn\\
[E_j^a(x_i), U(y_i, x_i)] &= - \delta_{ij} \delta^2({\bs 0}) U(y_i, x_i) T^a\,,
\end{align}
where $x_i$ and $y_i$ are the $i$-th components of ${\bs x}$ and ${\bs y}$ respectively. In our notation, the starting ${\bs x}$ and ending ${\bs y}$ points of the Wilson line only differ in the $i$-th component, which is reflected in the delta function $\delta^2({\bs 0})$ for the other components.

Now we construct the lattice Hamiltonian for Eq.~\eqref{sm:H_continuum}. The spatial lattice is set up as ${\bs x} = a(n_x, n_y,n_z) \equiv {\bs n} a$ where $n_i$'s are integers and $a$ is the spatial lattice spacing. We introduce the link variable
\begin{align}
U({\bs n}, \hat{i}) = \exp(ia A_i({\bs n})) = \exp(ia A_i^a({\bs n})T^a) \,,
\end{align}
which is the lattice version of the Wilson line with a straight line path from ${\bs n}$ to ${\bs n}+\hat{i}$. Here $\hat{i}$ denotes a unit vector along the $i$-th direction. The link variable $U({\bs n}, \hat{i})$ resides on the link from ${\bs n}$ to ${\bs n}+\hat{i}$ while $U^\dagger({\bs n}, \hat{i})$ lives on the link from ${\bs n}+\hat{i}$ to ${\bs n}$. The plaquette variable is defined as
\begin{align}
U_{ij}({\bs n}) = U^\dagger({\bs n},\hat{j}) U^\dagger({\bs n}+\hat{j},\hat{i}) U({\bs n}+\hat{i},\hat{j}) U({\bs n}, \hat{i}) \,,
\end{align}
where the multiplication is from right to left. We can show that
\begin{align}
U_{ij}({\bs n}) = \exp[ia^2F_{ij}({\bs n})+\ml{O}(a^3)] \,.
\end{align}
Therefore we can write the magnetic part of the Hamiltonian density as
\begin{align}
\frac{1}{4g^2}F_{ij}^aF_{ij}^a({\bs n}) = \frac{1}{a^4g^2}\sum_{i,j} \Big( \Tr \{ {\rm Re}[1-U_{ij}({\bs n})] \} +\ml{O}(a) \Big) = \frac{1}{a^4g^2} \sum_{i}\sum_{j>i} \Big( \Tr[2-U_{ij}({\bs n})-U^\dagger_{ij}({\bs n})] +\ml{O}(a) \Big)\,,
\end{align}
where the trace is over SU($N_c$) indexes. We will replace gauge fields with link variables in the lattice Hamiltonian. The next thing we need to work out is the commutation relation between electric fields and link variables. There are two things that we need to be careful. Firstly, according to Eq.~\eqref{sm:commutator_continuum}, the electric field can generate gauge transformations on both the left and right hand sides of the Wilson line. In the lattice formulation, if we want electric fields to live on links as link variables do, we need to introduce left ${\bs E}_L^a$ and right ${\bs E}_R^a$ electric fields that satisfy 
\begin{align}
\label{sm:commutator_lattice}
\big[ E_{Li}^a({\bs n}+\frac{\hat{i}}{2}), U({\bs n},\hat{j}) \big] &= -\delta_{ij} T^a  U({\bs n},\hat{j}) \,,\nn\\
\big[ E_{Ri}^a({\bs n}+\frac{\hat{i}}{2}), U({\bs n},\hat{j}) \big] &= -\delta_{ij}  U({\bs n},\hat{j}) T^a \,,
\end{align}
where the argument ${\bs n}+\frac{\hat{i}}{2}$ of the electric fields indicates that they live on the link between ${\bs n}$ and ${\bs n}+\hat{i}$. We note that the $i$-component of the electric field only lives on the link between ${\bs n}$ and ${\bs n}+\hat{i}$. The three components of the electric field at a lattice site now live on the three links that start at that lattice site and point along the spatial direction of each axis. Since these electric fields serve as generators of gauge transformation, they satisfy the following commutation relation as the generators of the SU($N_c$) group
\begin{align}
\label{sm:nonabelian}
[E_{Li}^a, E_{Li}^b] = if^{abc} E_{Li}^c\,,\quad [E_{Ri}^a, E_{Ri}^b] = -if^{abc} E_{Ri}^c \,,
\end{align}
where there is no summation over $i$. When summing over all links, we have
\begin{align}
\sum_{\rm links} (E_{Li}^a)^2 = \sum_{\rm links} (E_{Ri}^a)^2 \,.
\end{align}
So we can use either of them to represent the electric term in the Hamiltonian. The second thing that we need to modify is the mass dimension of the electric field. When imposing Eq.~\eqref{sm:commutator_lattice}, we implicitly change the mass dimension of the electric field. To see this explicitly, we can expand both sides of Eq.~\eqref{sm:commutator_lattice} to lowest non-trivial order, which gives
\begin{align}
[E_{Lj}^a, A_i^b] \approx \frac{i}{a} \delta_{ij}\delta^{ab} \,,\quad [E_{Rj}^a, A_i^b] \approx \frac{i}{a} \delta_{ij}\delta^{ab} \,.
\end{align}
Comparing this with the discretized version of Eq.~\eqref{sm:EA_continuum}, we see that a factor of $a^2$ has been absorbed into the definition of $E_{Lj}^a$ and $E_{Rj}^a$, i.e., $a^2E_j^a \to E_{Lj}^a, E_{Rj}^a$. Putting everything together, the lattice version of the Hamiltonian can be written as
\begin{align}
H = \int \diff^3 x\ml{H}({\bs x}) = a^3 \sum_{\bs n} \ml{H}({\bs n}) = \frac{g^2}{2a}\Big( \sum_{\rm links} ({\bs E}_{L}^a)^2 + \frac{2}{g^4} \sum_{\rm plaquettes} \sum_{i}\sum_{j>i} \Tr[2-U_{ij}-U^\dagger_{ij}] \Big)\,.
\end{align}
In the lattice Hamiltonian, it is equivalent to use ${\bs E}_{R}^a$ instead of ${\bs E}_{L}^a$. For $2+1$ dimensional gauge theory, the prefactors of the electric and magnetic terms are $g^2/2$ and $1/(a^2g^2)$ respectively.

Next we discuss how to impose the lattice version of the Gauss's law. Physically, the Gauss's law means physical states are gauge invariant. In the continuum, the Gauss's law requires that $D_iE_i$ vanishes when acting on physical states. We note that $D_iE_i$ is the generator of gauge transformation in the continuum, as shown in Eq.~\eqref{sm:DE_generator}.  However, in the lattice version, the generators are $E_{Li}^a$ and $E_{Ri}^a$. So to impose the Gauss's law in the lattice construction, we require at each spatial site ${\bs n}$, the following operator gives zero when acting on physical states:
\begin{align}
\label{sm:gauss}
\sum_{i} E_{Li}^a({\bs n}+\frac{1}{2}\hat{i}) - \sum_{i} E_{Ri}^a({\bs n}-\frac{1}{2}\hat{i}) \to 0\,,
\end{align}
where $E_{Li}^a$ lives on the link starting at ${\bs n}$ and ending at ${\bs n}+\h{i}$ while $E_{Ri}^a$ resides on the link starting at ${\bs n}-\h{i}$ and ending at ${\bs n}$.

Finally we comment on the sign convention. In practical calculations, it would be more convenient to flip the sign of $E_R^a$, i.e., $E_R^a\to -E_R^a$. As a result, Eqs.~(\ref{sm:commutator_lattice}),~\eqref{sm:nonabelian} and~\eqref{sm:gauss} become
\begin{align}
& \big[ E_{Li}^a({\bs n}+\frac{\hat{i}}{2}), U({\bs n},\hat{j}) \big] = -\delta_{ij} T^a  U({\bs n},\hat{j}) \,,\nn\\
& \big[ E_{Ri}^a({\bs n}+\frac{\hat{i}}{2}), U({\bs n},\hat{j}) \big] = \delta_{ij}  U({\bs n},\hat{j}) T^a \,,\nn\\
& [E_{Li}^a, E_{Li}^b] = if^{abc} E_{Li}^c\,,\quad [E_{Ri}^a, E_{Ri}^b] = if^{abc} E_{Ri}^c \,, \nn\\
& \sum_{i} E_{Li}^a({\bs n}+\frac{1}{2}\hat{i}) + \sum_{i} E_{Ri}^a({\bs n}-\frac{1}{2}\hat{i}) \to 0\,.
\end{align}

With these conventions, one can work out the matrix element of a plaquette operator $\square\equiv\Tr(U_{12})$ on a plaquette chain for physical states of 2+1 dimensional SU(2) lattice gauge theory~\cite{Klco:2019evd,PhysRevD.104.034501}
\begin{align}
\langle J_1J_2J_3J_4 | \square | j_1j_2j_3j_4 \rangle = \prod_{\beta=a,b,c,d}(-1)^{j_\beta}
\prod_{\alpha=1,2,3,4} \left[ (-1)^{j_\alpha+J_\alpha}\sqrt{(2j_\alpha+1)(2J_\alpha+1)} \right]  \nn \\
\left\{ \begin{array}{ccc}  j_a & j_1 & j_2 \\ 1/2 & J_2 & J_1   \end{array}  \right\} \left\{ \begin{array}{ccc}  j_b & j_2 & j_3 \\ 1/2 & J_3 & J_2   \end{array}  \right\}
\left\{ \begin{array}{ccc}  j_c & j_3 & j_4 \\ 1/2 & J_4 & J_3   \end{array}  \right\}
\left\{ \begin{array}{ccc}  j_d & j_4 & j_1 \\ 1/2 & J_1 & J_4  \end{array}  \right\} \,,
\end{align}
in which the Wigner 3-j symbols are used and the physical states are labeled as in Fig.~\ref{fig:square}. The plaquette operator acts on the links $1,2,3,4$. Here $j_1j_2j_3j_4$ specifies the initial state on the square with solid lines while $J_1J_2J_3J_4$ specifies the final state. At each vertex, the Gauss's law has been taken into account to construct a physical state that transforms as a SU(2) singlet.

\begin{figure}
\centering
\includegraphics[width=0.45\textwidth]{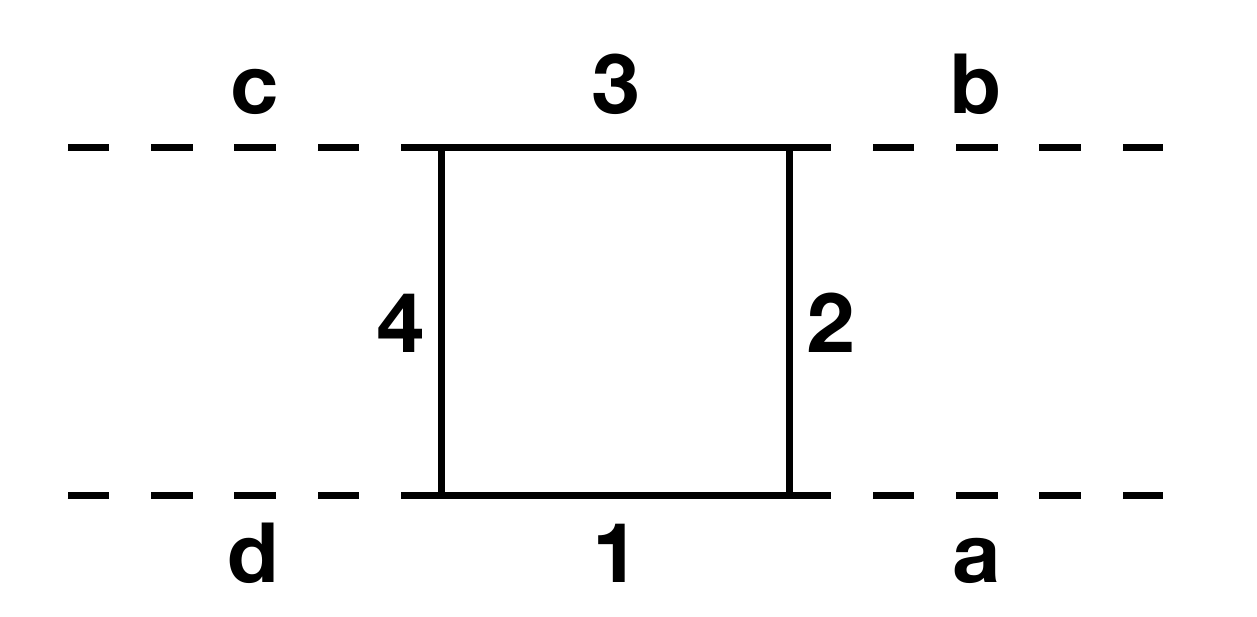}
\caption{A square plaquette on a plaquette chain. A plaquette operator that acts on the links $1,2,3,4$ is under consideration. The links $a,b,c,d$ are external and the states on them are not changed by the plaquette operator.}
\label{fig:square}
\end{figure}

\subsection{Momentum Basis for Spin Chain}
We consider a periodic spin chain with $N$ lattice sites. The Hamiltonian of the system is made up of
\begin{align}
H = H_{zz} + H_z + H_x \,,\quad H_{zz} = \sum_{i=0}^{N-1} \sigma_i^z\sigma_{i+1}^z \,,\quad H_{z} = \sum_{i=0}^{N-1} \sigma_i^z \,, \quad H_{x} = \sum_{i=0}^{N-1} \sigma_i^x \frac{1-3\sigma_{i-1}^z}{4}\frac{1-3\sigma_{i+1}^z}{4}\,.
\end{align}
Due to the periodic boundary condition, the Hamiltonian is invariant under translation. We define $\h{T}$ as the operator for a translation by one lattice site. The condition $[H, \h{T}]$ means we can simultaneously diagonalize $H$ and $\h{T}$. The eigenstates of the translation operator $\h{T}$ correspond to states in momentum space. To define them, we first construct equivalent classes under the translation $\h{T}$. Each equivalent class consists of states that are related via translations. For example, if we have
\begin{align}
|a_j\rangle = \h{T}^{n_{ji}} |a_i\rangle \,,
\end{align}
for some integer $n_{ji}\in\{1,\cdots,N-1\}$, then the two states $|a_i\rangle$ and $|a_j\rangle$ are in the same equivalent class. For each equivalent class, we choose one representative state $|a\rangle$. Then the momentum states that are based on $|a\rangle$ are defined by
\begin{align}
|a(k)\rangle = \frac{1}{\sqrt{N_a}} \sum_{r=0}^{N-1} e^{-ikr} \hat{T}^r |a\rangle \,,
\quad {\rm for}\ k = \frac{2\pi}{N} n_k \,, \ n_k\in\{0,1,\cdots,N-1\} \,.
\end{align}
They are the eigenstates of $\h{T}$ with the eigenvalues $e^{-ik}$. Not all momentum states written as above exist. The existence condition depends on the periodicity of the representative state $|a\rangle$, which is defined as the smallest nonzero integer $R_a$ such that
\begin{align}
\hat{T}^{R_a} |a\rangle = |a\rangle \,.
\end{align}
A momentum state $|a(k)\rangle$ is a valid state if and only if $kR_a$ is a multiple of $2\pi$. If a momentum state $|a(k)\rangle$ exists, its normalization factor is given by
\begin{align}
N_a = \frac{N^2}{R_a} \,.
\end{align}

After introducing the momentum basis, we write down the matrix elements of the Hamiltonian and the operators studied in the main text. First for $H_{zz}$ and $H_z$, we have
\begin{align}
\langle b(k') | H_{zz} | a(k) \rangle & = \delta_{ab}\delta_{k'k} \sum_{i=0}^{N-1} z_i(a)z_{i+1}(a) \,,\nn\\
\langle b(k') | H_{z} | a(k) \rangle & = \delta_{ab}\delta_{k'k} \sum_{i=0}^{N-1} z_i(a) \,,
\end{align}
where $z_i(a)$ is given by $\sigma_i^z|a\rangle = z_i(a)|a\rangle$ and it is $+1$ ($-1$) if the $i$-th lattice of $|a\rangle$ is spin-up (spin-down). We note that both $H_{zz}$ and $H_z$ are diagonal. Then we write down the matrix for $H_x$:
\begin{align}
\langle b(k') | H_x | a(k) \rangle = \delta_{k'k} \sqrt{\frac{N_b}{N_a}} \sum_{i=0}^{N-1} e^{-ik\ell_i} \frac{1-3z_{i-1}(a)}{4} \frac{1-3z_{i+1}(a)}{4} \,,
\end{align}
where $\ell_i$ is an integer that depends on $i$ and determined by
\begin{align}
\sigma_i^x |a\rangle = \hat{T}^{-\ell_i} |b\rangle \,. 
\end{align}
Finally, we write down the matrix elements for the 1-plaquette ($O_1$) and 2-plaquette ($O_2$) operators. Without loss of generality, we can assume they sit at the site $i=0$, i.e.,
\begin{align}
O_1(i=0) = \sigma_0^x \frac{1-3\sigma_{-1}^z}{4} \frac{1-3\sigma_{1}^z}{4} \,,\qquad O_2(i=0) = \sigma_0^x\sigma_1^x \frac{1-3\sigma_{-1}^z}{4} \frac{1-3\sigma_{2}^z}{4} \frac{1+3\sigma_0^z\sigma_1^z}{4} \,.
\end{align}
After some algebra, we find
\begin{align}
\langle b(k') | O_1(i=0) | a(k) \rangle = \frac{1}{N} \sqrt{\frac{N_b}{N_a}} \sum_{r=0}^{N-1} e^{i(k'-k)r-ik'\ell_{r}} \frac{1-3z_{-1-r}(a)}{4} \frac{1-3z_{1-r}(a)}{4} \,,
\end{align}
where $\ell_{r}$ is an integer that depends on $r$ and determined by
\begin{align}
\sigma^x_{-r} |a\rangle = \hat{T}^{-\ell_{r}}|b\rangle \,.
\end{align}
Similarly, we have
\begin{align}
\langle b(k') | O_2(i=0) | a(k) \rangle = \frac{1}{N} \sqrt{\frac{N_b}{N_a}} \sum_{r=0}^{N-1} e^{i(k'-k)r-ik'\ell_{r}} \frac{1-3z_{-1-r}(a)}{4} \frac{1-3z_{2-r}(a)}{4} \frac{1+3z_{-r}(a)z_{1-r}(a)}{4} \,,
\end{align}
where $\ell_{r}$ depends on $r$ and is given by
\begin{align}
\sigma^x_{-r}\sigma^x_{1-r} |a\rangle = \hat{T}^{-\ell_{r}}|b\rangle \,.
\end{align}

\begin{figure}
\centering
\includegraphics[width=0.45\textwidth]{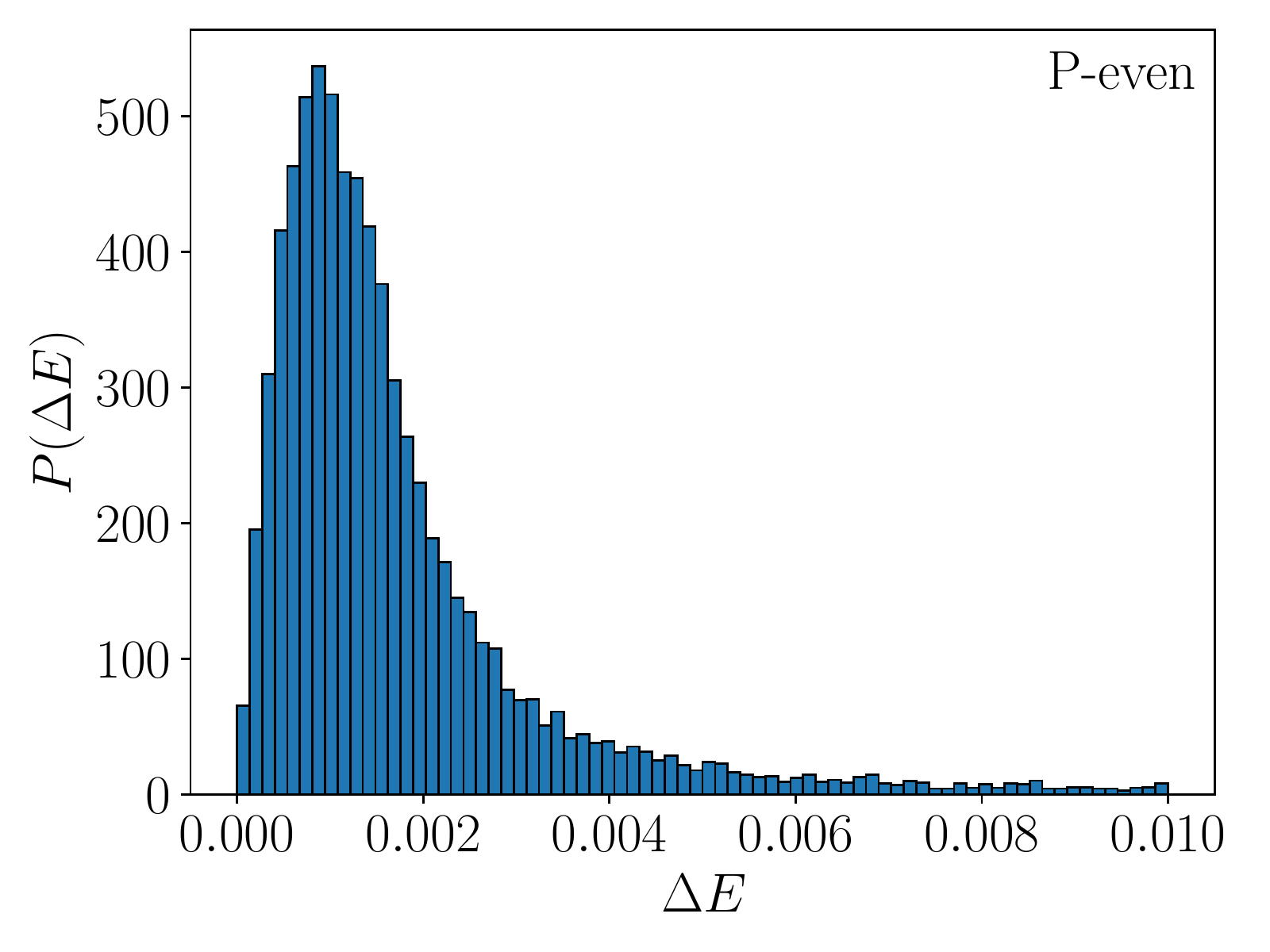}
\includegraphics[width=0.45\textwidth]{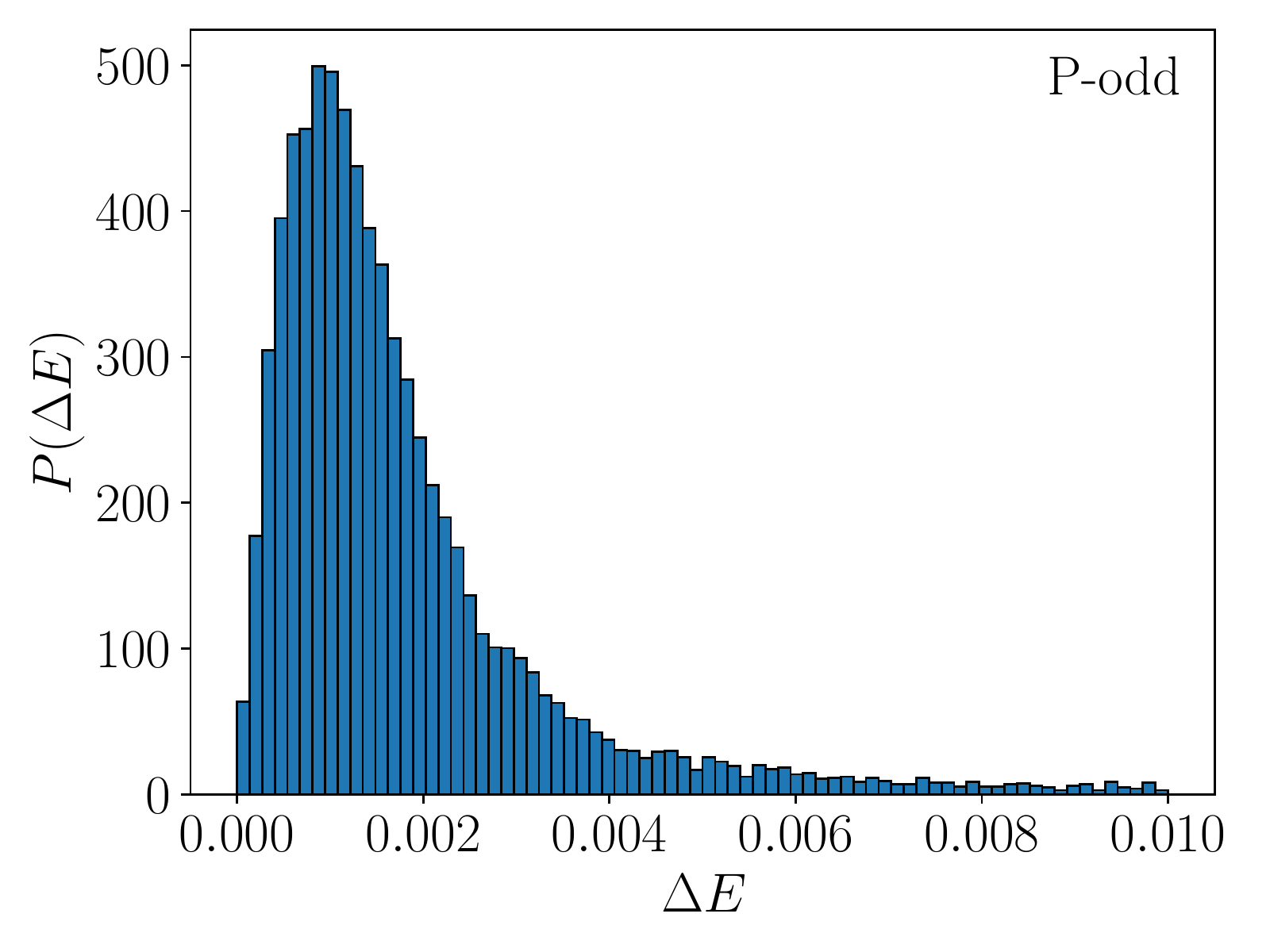}
\caption{Distributions of energy level spacing $\Delta E$ in the even (left) and odd (right) parity sectors of the $k=0$ sector for $N=19$, which exhibit level repulsion in each case.}
\label{fig:wd_parity}
\end{figure}

\begin{figure}
\centering
\includegraphics[width=0.45\textwidth]{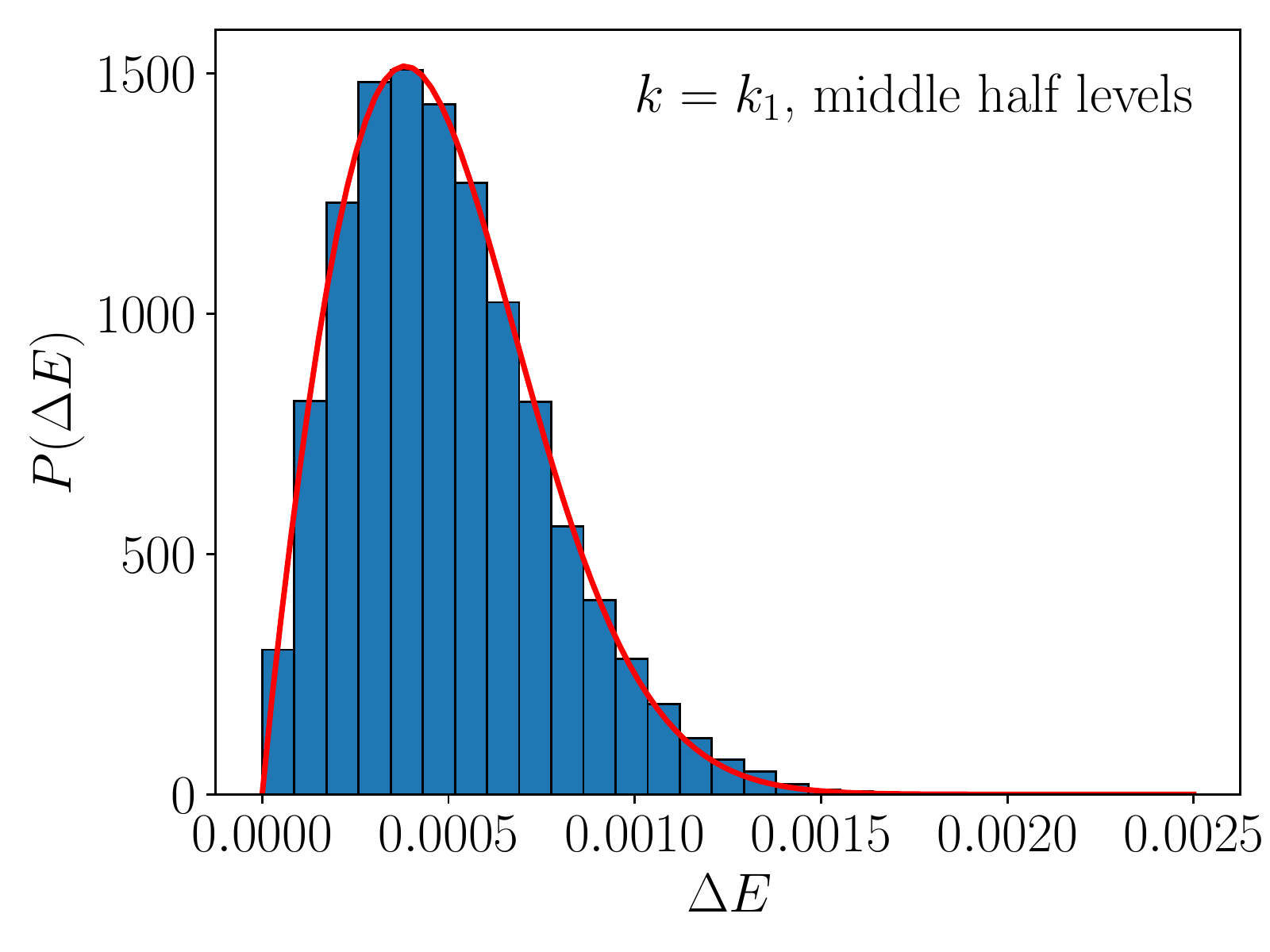}
\caption{Distribution of energy level spacing in the middle half of the spectrum for the $k_1$ sector in the case of $N=19$.}
\label{fig:wd_half}
\end{figure}

\subsection{Additional Plots and Fitting Parameter Values}

To really see the effect of the remaining parity symmetry on the level statistics in the $k=0$ sector, we further separate the $k=0$ sector into a parity-even sector and a parity-odd sector. We then calculate the nearest eigenenergy gaps in each parity sector and plot their distributions for the $N=19$ case in Fig.~\ref{fig:wd_parity}. The distribution of energy level spacing in each parity sector exhibits level repulsion.

In Fig.~\ref{fig:wd}, the Poisson function and the Wigner surmise are used to fit the distributions of energy level spacing in four momentum sectors. For the Poisson function fit in the $k=0$ sector, the fitted parameter values are $a=1231.93$ and $b=1108.64$. For the Wigner surmise fits in the other three momentum sectors, the fitted parameter values are $a=5.62\times10^5$, $b=7.54\times 10^{-1}$, $c=1.96\times10^6$ for the $k_1$ sector, $a=4.51\times10^5$, $b=7.27\times 10^{-1}$, $c=1.92\times10^6$ for the $k_2$ sector and $a=6.54\times10^5$, $b=7.72\times 10^{-1}$, $c=2.00\times10^6$ for the $k_3$ sector. The fit in the tail region can be improved by focusing on the energy levels in the middle of the spectrum, by removing those levels near the lower and upper ends of the spectrum. For example, when we focus on the middle half of the energy levels in the $k_1$ sector in the case of $N=19$, the distribution of energy level spacing is better described by the Wigner surmise, as shown in Fig.~\ref{fig:wd_half}. The fitted parameter values are $a=3.48\times10^5$, $b=9.25\times 10^{-1}$, $c=3.15\times10^6$.

Fig.~\ref{fig:diag} in the main text seems to suggest the difference between the diagonal matrix element and the microcanonical ensemble proxy decays faster than an exponential in $N$. To investigate this, we change the $x$-axis from $N$ to $N^2$ and the result is shown on the left of Fig.~\ref{fig:diag_N2}. The lines decrease almost linearly, which indicates the difference decreases exponentially in $N^2$ rather than $N$. To better understand this, we use only the middle two thirds of the eigenstates (ordered by their eigenenergies) in the calculation of the average and depict the results on the right of Fig.~\ref{fig:diag_N2}. The last four points exhibit a better agreement with an exponential decrease in $N$.

\begin{figure}
\centering
\includegraphics[width=0.45\textwidth]{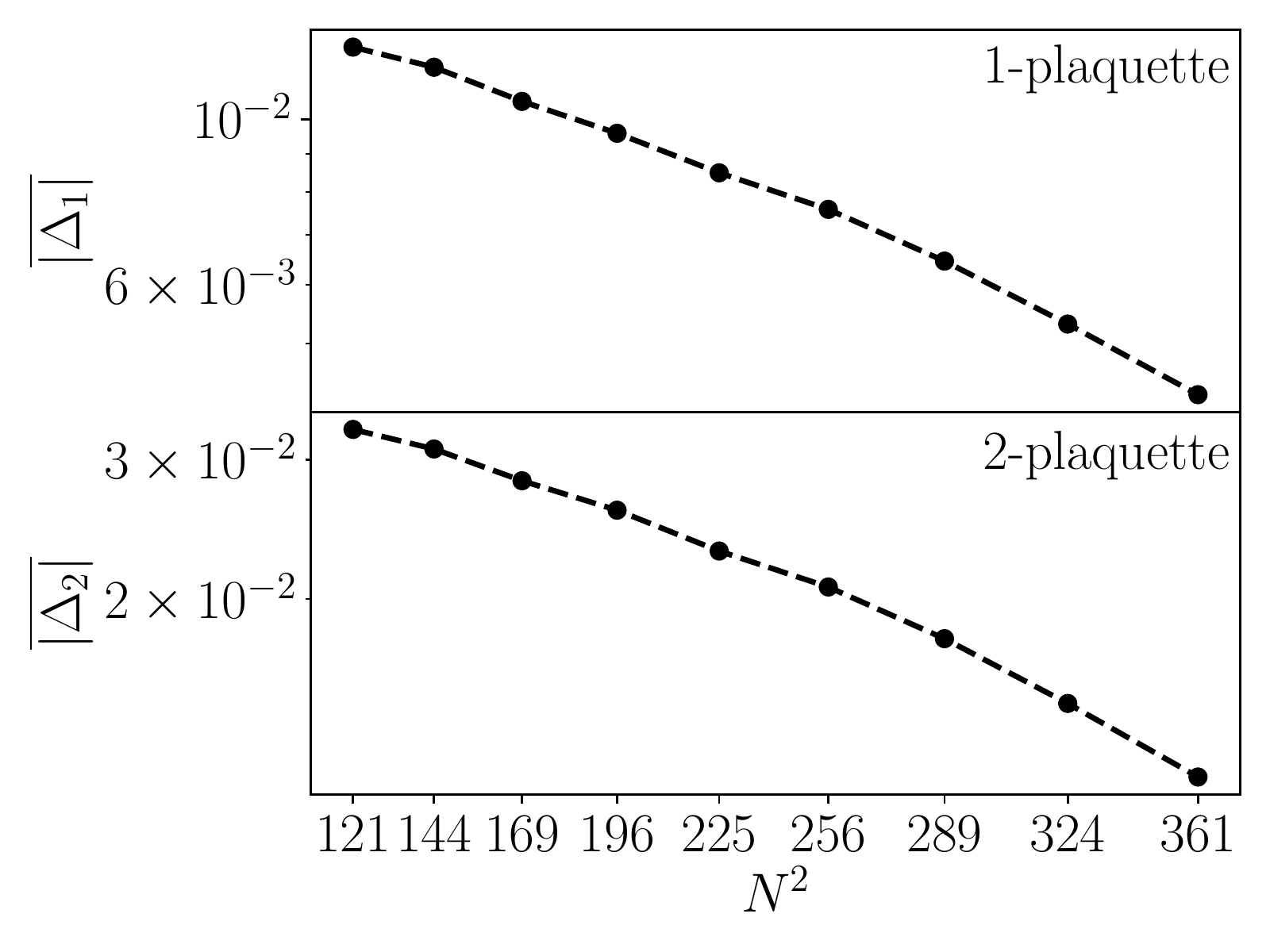}
~
\includegraphics[width=0.45\textwidth]{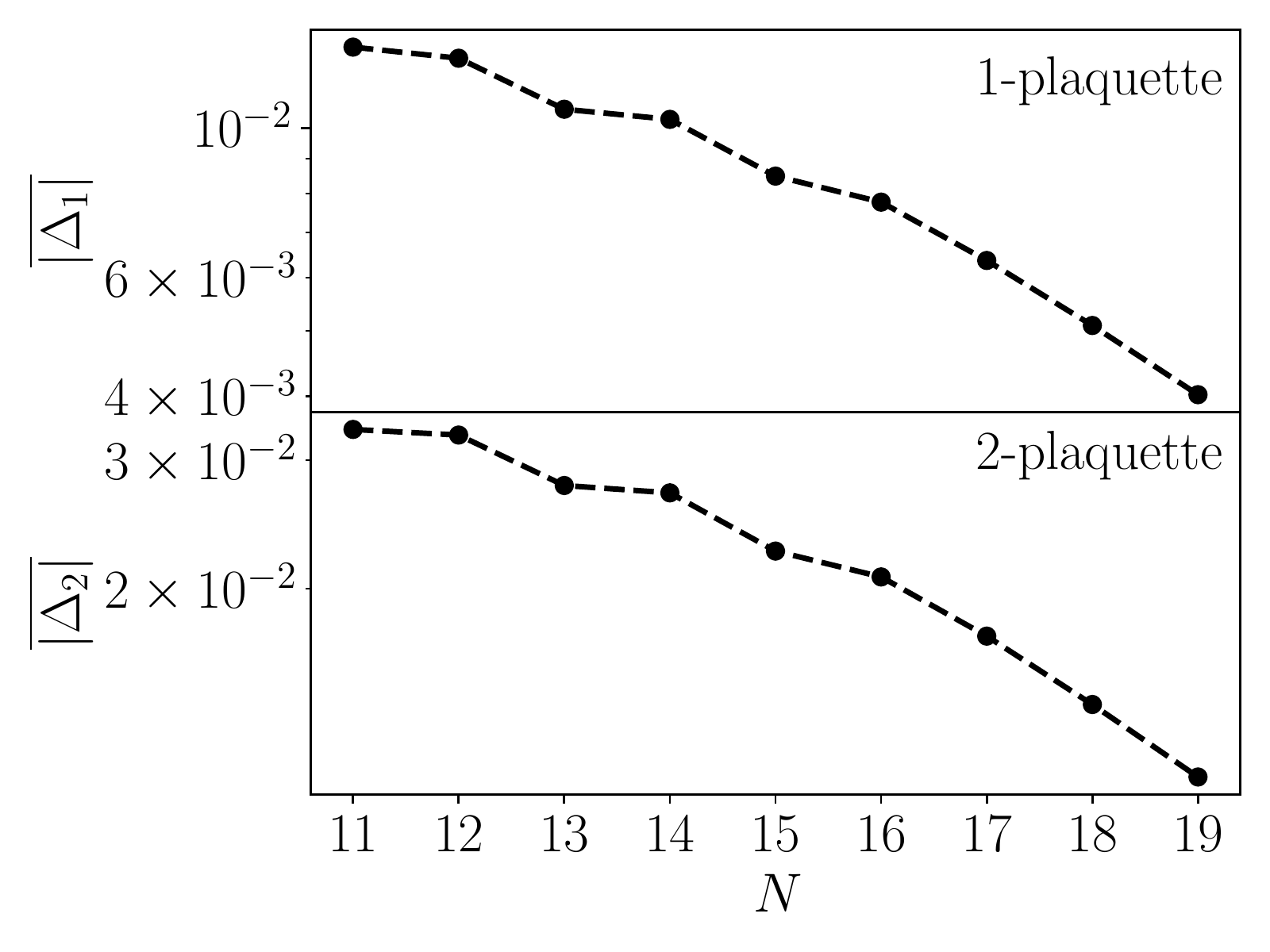}
\caption{Left: same figure as Fig.~\ref{fig:diag} except for the $x$-axis, which is $N^2$ here. Right: same as Fig.~\ref{fig:diag} but only the middle two thirds of the eigenstates are used in calculating the average.}
\label{fig:diag_N2}
\end{figure}

\end{widetext}

\end{document}